\documentclass[preprint,aps]{revtex4}
\usepackage{graphicx}% Include figure files
\usepackage{dcolumn}% Align table columns on decimal point
\usepackage{bm}% bold math
\usepackage{subfigure}

\begin{document}

\title{Transient behavior of the thermocapillary migration of drops under the influence of deformation}

\author{L. Chang}
\affiliation{National Microgravity Laboratory, Institute of Mechanics, Chinese Academy of Sciences, Beijing 100190, P.R.China}

\author{Z. Yin}
\email{zhaohua.yin@imech.ac.cn}
\affiliation{National Microgravity Laboratory, Institute of Mechanics, Chinese Academy of Sciences, Beijing 100190, P.R.China}

\author{W. Hu}
\affiliation{National Microgravity Laboratory, Institute of Mechanics, Chinese Academy of Sciences, Beijing 100190, P.R.China}

\date{\today}

\begin{abstract}
The transient thermocapillary migration of drops with nontrivial deformation is studied. The finite
difference method is employed to solve the incompressible Navier-Stokes equations coupled with the energy equation; the front-tracking method is adopted to track the moving deformable drop interface. In the hot region, deformations of drops increase with the decrease of interfacial tensions. In order to indicate the temperature impact on the interfacial tension, a local capillary number ($Ca_l$) is introduced. It is found that, when the drop density is smaller/larger than that of the bulk fluid, the drop velocity decreases/increases with the increase of the drop deformation.
\end{abstract}


\maketitle

\section{Introduction}

Thermocapillary migration, which is induced by the variance of the interfacial tension when a temperature
gradient is imposed on the bulk fluid, is an important phenomenon in drop/bubble transportation. With the
development of aeronautics and astronautics, it is necessary to study the thermocapillary motion because of its practical roles in the material processing, and the management of heat and fluids in space \cite{Ratke 2000, Ostrach 1982}. In most studies on the thermocapillary motion, it is assumed that the interfacial tension linearly depends on the temperature:
\begin{equation}
\sigma(T)=\sigma_0+\sigma_T(T-T_0),\label{eq1}
\end{equation}
where the constant $\sigma_T$ is the interfacial tension coefficient of temperature, and $\sigma_0$ the interfacial tension at the reference temperature $T_0$.

Considering the balance between the capillary force and viscosity on the bubble or drop, the reference velocity can be defined as
\[U=|\sigma_T| |\nabla\ T_\infty| R_0/\mu_1,\]
in which $|\nabla\ T_\infty|$ is the temperature gradient imposed on bulk fluid (the subscript $1$ represented), $\mu$ the viscosity, and $R_0$ the radius of the spherical drop/bubble.

\begin{table*}[h]
\begin{center}
\begin{tabular}{c|c|c}\toprule
&$\sigma_0(10^{-3}N/m)$& $\sigma_T (10^{-3} N/(m \cdot K))$ \\ \hline
Gas bubble/Silicone oil\cite{Hadland 1999} &17.3&-0.061  \\ \hline
Fc-75 drop/Silicone oil\cite{Hadland 1999} &3.47&-0.036 \\ \hline
Monotectic alloy\cite{Hoyer 2006} &5.7&-0.23\\ \hline \hline
\end{tabular}
\caption{\label{tab:W1W2W} Typical interfacial tensions and their temperature coefficients.}
\end{center}
\end{table*}

The capillary number is defined as
\[
Ca=\mu_1 U/\sigma_0=|\sigma_T| |\nabla T_\infty| R_0/\sigma_0.
\]
In most cases, the $Ca$ numbers in thermocapillary migration are small (0.01 or less, e.g., the materials in the first two lines of Table 1), and the deformations of drops are also very small. Therefore, previous studies usually assumed drops to be nondeformable \cite{Yin 2008}. However, in practice, there are many situations where the interface tension is very small, or where the deformation introduced by capillary effect is very large. For the case of the preparation of monotectic alloys in micro-gravity conditions \cite{Carlberg 1980}, the $Ca$ number of the Al-based alloy $(Al_{34.5}Bi_{65.5})_{95} Sn_{5}$ is 0.12 at the temperature of 1200k if $R_0 = 300\mu m$ and $|\nabla T_\infty| = 100K/cm$ \cite{Li 2008} (see the third line in Table 1). Such a big $Ca$ number can cause obvious deformation on the drop. Similar situations happen in the chemical flooding methods applied in enhanced oil recovery \cite{Gurgel 2008,Xia 2006, Lyford 1998a, Lyford 1998b, Morrow 2001}.
Finally, experiments in space also observed some slight deformation of fairly large bubbles \cite{Hadland 1999}. Therefore, it is necessary to investigate the deformed drops with relatively large $Ca$'s or small interfacial tensions in the thermocapillary studies.

When inertia and thermal convections are neglected, the velocity of the spherical drop in thermocapillary motion is presented in the pioneering work of Young \emph{et al. }\cite{Young 1959}:
\[V_{YGB}=2AU,\]
where
\begin{equation}
A=\frac{1}{(2+3\alpha)(2+\lambda)}, \label{eq2}
\end{equation}
and $\alpha=\mu_2/\mu_1$ and $\lambda=k_2/k_1$ are the ratios of the kinematic viscosity and thermal
conductivity between droplet (the subscript $2$ represented) and background liquid, respectively.

There are a lot of investigations studying the thermocapillary motion, which are reviewed in \cite{SubBala 2001, SubBalaWoz 2002}. Most of the researches sofar focus on the convection effects of inertia and energy, and only a few of them are about the drop/bubble deformation. When the thermal convection
is very small, the exact solution of the momentum equation was provided in \cite{Balasubramaniam
1987}, where small inertial deformations of drops were also calculated. Under the same assumption,
using the Lorentz reciprocal theorem, Haj-Hariri \emph{et al.} calculated the small inertial
deformations as well as the consequent correction on the temperature field and the migration velocity \cite{Haj-Hariri 1990}. Later, the work is extended to moderate parameters in a steady-state numerical investigation \cite{Haj-Hariri 1997}. It is concluded that even a small deformation will retard the motion of the drop, and the steady-state migration can still be reached even for large capillary numbers up to 0.5. However, the actual continuous decrease of the interfacial tension when the drop migrates to the hot region was not considered in their study. While taking this decrease into consideration, bubbles with large capillary numbers can not reach steady migrating states \cite{Welch 1998}.

According to our knowledge, transient migration behaviors due to drop deformation have never been reported before, which is the main topic in this work. This paper is arranged as follows: the governing equations and numerical methods are introduced in Section $2$, the validation tests are
presented in Section $3$, and the results and discussions are in Section $4$.

\begin{figure}
\begin{minipage}[t]{0.25\linewidth}
\scalebox{1}[1]{\includegraphics[width=\linewidth]{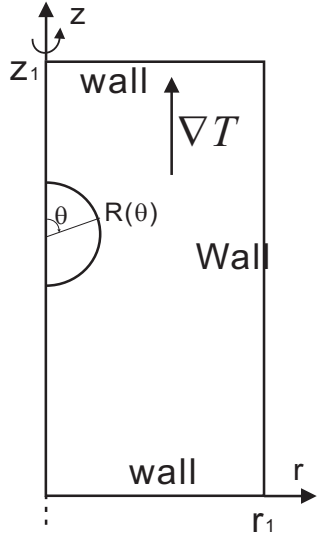}}
\end{minipage}
\caption{\label{fig:iso_time_distance_Ma}
The sketch of thermocapillary motion in the axisymmetric model.}
\end{figure}

\section{Governing equations and numerical methods}

In our axisymmetric simulation, the initially-spherical droplet with a radius of $R_0$ is surrounded by the bulk liquid, which is imposed on a constant temperature gradient along the $z$ axis (Fig. 1). The calculated domain is $\Omega((r, z))$ ($r \in[0, r_1]$, $z \in[0, z_1]$), and the center of the droplet is at $r=0$. The governing equations for the thermocapillary migration of droplets are:
\begin{equation}
 \nabla\cdot \textbf{u}=0,\label{eq3}
\end{equation}
\begin{equation}
  \frac{\partial {(\rho \textbf{u})}}{\partial
{t}}+\nabla\cdot(\rho \textbf{u} \textbf{u})=-\nabla
p+\nabla\cdot(\mu(\nabla \textbf{u}+\nabla^T
  \textbf{u}))+\textbf{F}_{\sigma}, \label{eq4}
\end{equation}
\begin{equation}
 \rho C_p(\frac{\partial T}{\partial {t}}+\textbf{u}\cdot\nabla
  T)=\nabla\cdot(\emph{k}\nabla T), \label{eq5}
\end{equation}
in which $\textbf{u}=(v_r,v_z)$ is the velocity, $\rho$ the density, $p$ the pressure , and $C_p$ the specific heat. $\emph{\textbf{F}}_{\sigma}$ is the body force term due to the interfacial tension, and its axisymmetrical formula is presented in \cite{Unverdi 1992a, Tryggvason 2001}:
\begin{equation}
\textbf{F}_{\sigma}=\int_{B} \delta(\textbf{x}-\textbf{x}_{f})(\sigma \kappa
\textbf{n}+\frac{\partial \sigma}{\partial s}\bm{\tau})d s. \label{eq6}
\end{equation}
Here, $\textbf{x}=(r,z)$ is the space vector, $\textbf{x}_f=(r_f,z_f)$ the position of the cell $f$ on the interface $B$, and $s$ the natural coordinate along the interface. $\textbf{n}=(n_r,n_z)$ and $\bm{\tau}$ denote the normal and tangential unit vectors of the interface, respectively. $\kappa=\kappa_1+\kappa_2$ is the sum of two principal curvatures of the interface, $\kappa_1$ is the in-plane curvature and $\kappa_2$ is given
as \cite{Kreyszig 1991}:
\begin{equation}
\kappa_2=-\frac{n_r}{r_f}.  \label{eq7}
\end{equation}

We defined nondimensional variables as follows:
\[
 \bar{\textbf{u}}=\textbf{u}/U,\quad \bar{\textbf{x}}=\textbf{x}/R_0, \quad \bar{t}=t/(\frac{R_0}{U}), \quad \bar{p}=p/(\rho_1 U^2),
\]
\begin{equation}
 \bar{T}=T/(|\nabla T_{\infty}|R_0),\quad\bar{\rho}=\rho/\rho_1, \quad \xi=\rho_2/\rho_1, \quad\bar{\mu}=\mu/\mu_1, \label{eq8}
\end{equation}
\[
\bar{k}=k/k_{1},\quad\bar{C_p}=C_p/C_{p1}, \quad\gamma=C_{p2}/C_{p1},
\]
\[
\bar{\textbf{F}}_{\sigma}={\textbf{F}_{\sigma}}R_0/(\rho_1U^2),\quad Ma=\rho_1 C_{p1}UR_0/k_1.
\]

In total, the problem is governed by seven nondimensional parameters: $Re, Ma, Ca, \alpha, \lambda, \gamma$ and $\xi$. The nondimensional equations in the axisymmetrical model are:
\begin{equation}
\frac{\partial {\bar{v}_r}}{\partial {\bar{r}}}+\frac{\bar{v}_r}{\bar{r}}+\frac{\partial
{\bar{v}_z}}{\partial {\bar{z}}}=0,\label{eq9}
\end{equation}
\begin{eqnarray}
\frac{\partial {(\bar{\rho} \bar{v}_r})}{\partial
{\bar{t}}}+\frac{1}{\bar{r}}\frac{\partial {(\bar{r}\bar{\rho
}\bar{v}_r \bar{v}_r)}}{\partial {\bar{r}}}+\frac{\partial
{(\bar{\rho} \bar{v}_z \bar{v}_r)}}{\partial
{\bar{z}}}
 = -\frac{\partial {\bar{p}}}{\partial {\bar{r}}}
+\frac{1}{Re}\left[\frac{2}{\bar{r}}\frac{\partial}{\partial
{\bar{r}}}(\bar{r}\bar{\mu}(\frac{\partial
{\bar{v}_r}}{\partial{\bar{r}}})) +\frac{\partial}{\partial
{\bar{z}}}(\bar{\mu}(\frac{\partial
{\bar{v}_r}}{\partial{\bar{z}}}+\frac{\partial
{\bar{v}_z}}{\partial{\bar{r}}}))
- \frac{2\bar{\mu} \bar{v}_r}{\bar{r}^2}\right]
\nonumber \\ +\bar{F}_{\sigma r}, \label{eq10}
\end{eqnarray}
\begin{eqnarray}
\frac{\partial {(\bar{\rho} \bar{v}_z})}{\partial
{\bar{t}}}+\frac{1}{\bar{r}}\frac{\partial {(\bar{r}\bar{\rho}
\bar{v}_r \bar{v}_z)}}{\partial {\bar{r}}}+\frac{\partial
{(\bar{\rho }\bar{v}_z \bar{v}_z)}}{\partial
{\bar{z}}}=-\frac{\partial {\bar{p}}}{\partial
{\bar{z}}}
+\frac{1}{Re}\left[\frac{1}{\bar{r}}\frac{\partial}{\partial
{\bar{r}}}(\bar{r}\bar{\mu}(\frac{\partial
{\bar{v}_z}}{\partial{\bar{r}}}+\frac{\partial
{\bar{v}_r}}{\partial{\bar{z}}})) +2\frac{\partial}{\partial
{\bar{z}}}(\bar{\mu}(\frac{\partial
{\bar{v}_z}}{\partial{\bar{z}}}))\right] \nonumber\\+\bar{F}_{\sigma z}, \label{eq11}
\end{eqnarray}
\begin{equation}
\bar{\rho} \bar{C}_p \left( \frac{\partial \bar{T}}{\partial
{\bar{t}}}+\bar{v}_r \frac{\partial \bar{T}}{\partial
{\bar{r}}}+\bar{v}_z\frac{\partial \bar{T}}{\partial {\bar{z}}}
\right) =
 \frac{1}{Ma}\left[\frac{1}{\bar{r}}\frac{\partial}{\partial{\bar{r}}}(\bar{r}\bar{k}\frac{\partial{\bar{T}}}{\partial{\bar{r}}})
+\frac{\partial}{\partial{\bar{z}}}(\bar{k}\frac{\partial{\bar{T}}}{\partial{\bar{z}}})\right]. \label{eq12}
\end{equation}
Using Eq. \ref{eq1}, the nondimensional interfacial tension term
$\bar{\textbf{F}}_{\sigma}$ can be written as:
\begin{equation}
\bar{\textbf{F}}_{\sigma}=\int_{B}
\delta(\bar{\textbf{x}}-\bar{\textbf{x}}_{f})((\frac{1}{ReCa}-\frac{1}{Re}(\bar{T}-\bar{T}_0))
\kappa \textbf{n}-\frac{1}{Re}\frac{\partial \bar{T}}{\partial \bar{s}}\bm{\tau})d \bar{s}. \label{eq13}
\end{equation}

Eqs. \ref{eq9}-\ref{eq12} are valid for both phases of the drop and the bulk fluid. The discontinuities of the physical parameters across the interface are handled with an indicator function (see \cite{Hua 2007} for more details). The governing equations are discretized by the second-order center-difference method on fixed, staggered, Cartesian grids. The projection method \cite{Chorin 1967} is adopted to solve the Navier-Stokes equations.

On the symmetric axis, the symmetrical condition is employed:
\begin{equation}
\bar{v}_r|_{\bar{r}=0}=0,\quad \frac{\partial \bar{v}_z}{\partial \bar{r}}|_{\bar{r}=0}=0,\quad
\frac{\partial \bar{T}}{\partial \bar{r}}|_{\bar{r}=0}=0, \label{eq14}
\end{equation}
and all other boundaries for velocities adopt rigid boundary conditions, and those for temperature use constant temperature boundary conditions. Following the tradition in this field, the initial conditions are:
\begin{equation}
\bar{v}_r|_{\bar{t}=0}=\bar{v}_z|_{\bar{t}=0}=0,\quad \bar{T}|_{\bar{t}=0}=\bar{z}. \label{eq15}
\end{equation}
In the following, symbols without bars will be used to denote nondimensional values.

In addition, when the momentum equation is solved at each new time step, the front element $f$ is supposed to have the same velocity ($\textbf{u}_{f}$) as that of the surrounding fluid. In order to update the
front with a conserved drop volume, a velocity correction $\Delta u_{n}$ is imposed on $\textbf{u}_{f}$ in the normal direction ($\textbf{n}_f$) of the element:
\begin{equation}
\Delta u_{n}=-\frac{\sum_{f}\Delta s_{f}\textbf{u}_{f}\cdot \textbf{n}_{f}}{\sum_{f}\Delta s_{f}}. \label{eq16}
\end{equation}
With this correction, the drop volume is almost conserved throughout our simulations with a maximum
volume loss of less than 0.1\% of the initial volume. By comparison, without the velocity correction, there will be about 2\% loss at the simulating time $t=20$.

The computing domain in this paper is $(r,z) \in [0,6]\times[0,24]$, and the resolution is $128\times512$. The time step is $2\times10^{-4}$ for most cases, and $1\times10^{-4}$ for the $\xi=0.01$ case in subsection 4.2.

\section{The validation of the computing program}

\begin{figure}
\begin{minipage}[t]{0.45\linewidth}
\scalebox{1}[1]{\includegraphics[width=\linewidth]{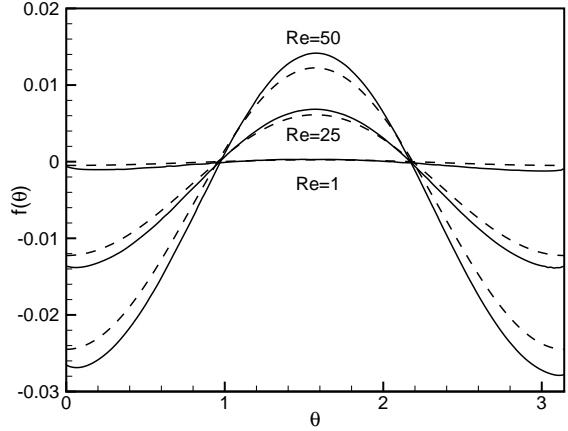}}
\end{minipage}
\caption{\label{fig:iso_time_distance_Ma}
Deviation of the drop profile from the sphere for $Ma=1$, $Ca=0.1$, $\alpha=\lambda=\xi=0.5$, $\gamma=0.25$, and $Re$ as indicated.  Dash line: the theoretical prediction in \cite{Haj-Hariri 1990}; solid line: the present work. }
\end{figure}

\begin{figure}
\begin{minipage}[t]{0.45\linewidth}
\scalebox{1}[1]{\includegraphics[width=\linewidth]{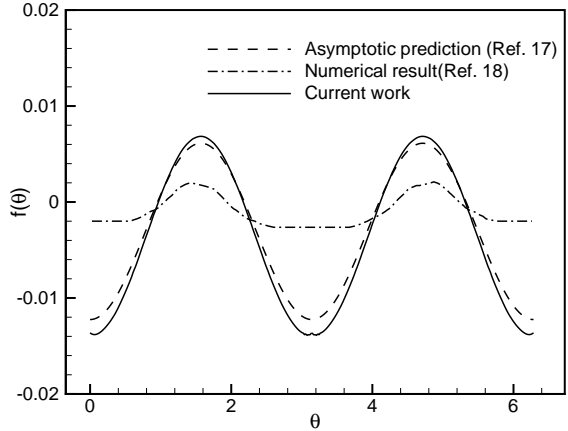}}
\end{minipage}
\caption{\label{fig:iso_time_distance_Ma}
Deviations of drop profiles from spheres in \cite{Haj-Hariri 1990}, \cite{Haj-Hariri 1997}, and this work, where $Re=25$, $Ma=1$, $Ca=0.1$, $\alpha=\lambda=\xi=0.5$, and $\gamma=0.25$.}
\end{figure}

In tradition, the scaled deviation of the drop profile from the sphere is defined as
\[f(\theta)=R(\theta)/R_0-1,\]
where, the polar angle $\theta$ is measured from the front stagnation point and $R(\theta)$ denotes the distance between the interface and the mass center of the drop (Fig. 1).

Before this work, there was a theoretical result of small inertial deformations of drops with small Ca and Re numbers \cite{Haj-Hariri 1990}:
\begin{equation}
f(\theta)=\frac{3}{8}A^2 Re Ca (\xi-1)(3\cos^2\theta-1). \label{eq17}
\end{equation}

The earlier numerical simulation \cite{Haj-Hariri 1997}, however, used a different formula for the interfacial tension:
\begin{equation}
\emph{\textbf{F}}_{\sigma}=\int_{B} \delta(\textbf{x}-\textbf{x}_{f})(\frac{1}{ReCa} \kappa
\textbf{n}-\frac{1}{Re}\frac{\partial T}{\partial s}\bm{\tau})d s. \label{eq18}
\end{equation}
Compared with Eq. \ref{eq13}, the effect of capillary force in Eq. \ref{eq18} was assumed constant on the normal direction of the interface, or, there is no decrease in the interfacial tension when the droplet moves to the hotter region \cite{Haj-Hariri 1997}. There are two distinguished differences between these two assumptions:
\begin{enumerate}
\item As it will be shown later in this paper, Eq. \ref{eq13} leads to a bigger drop deformation than Eq. \ref{eq18}.

\item With Eq. \ref{eq13}, the simulation must be stopped before the drop moves a distance of $(1/Ca-1)$ in the $z$ direction. This is because the interfacial tension of the drop will be smaller than zero beyond that location, and not only the simulation will collapse, but also it is not physically possible.

    With Eq. \ref{eq18}, on the other hand, the simulation can be extended to the infinity.
\end{enumerate}

In this section, Eq. \ref{eq18} is adopted in our codes to have some comparisons with the previous researches. It seems that our simulations have very good agreements with the analytical work (Fig. 2), and that a better similarity is achieved when the Reynolds number is smaller. Compared with the previous numerical work \cite{Haj-Hariri 1997}, our numerical simulations are closer to the theoretical analysis (Fig. 3).

In the rest of this paper, the interfacial tension is calculated by Eq. \ref{eq13}, and simulations are
stopped before the capillary force on the drop becomes negative.

\section{Results and discussions}
\subsection{Transient thermocapillary migrations of droplets with large capillary numbers}

In this subsection, we simulate the thermocapillary process with the same set of parameters in
\cite{Haj-Hariri 1997}: $Re=Ma=50$, $Ca=0.1$, $\alpha=\lambda=\xi=0.5$, and $\gamma=0.25$. The time evolution of migration velocity is shown in Fig. 4 (the solid line). When the capillary number is sufficiently small ($Ca=0.05$, the dashed line in Fig. 4), the drop will reach a steady migrating state after the initial accelerating process. However, the velocity with $Ca=0.1$ has an accelerating decrease when $t>60$, and can not reach a steady state. This result is different from that of \cite{Haj-Hariri 1997}, in which the steady velocity state was assumed in advance.

The deviation of the drop profile from the sphere for $Ca=0.1$ at different times is plotted in Fig. 5. It seems that when the drop migrates to the hotter region, the deformation keeps increasing and the drop loses its fore and aft symmetry. At the late stage, the bottom of the drop becomes flattened, and the drop looks like a cap (Fig. 6(b)). The deformation at $t=60$ in the current study is about twice bigger than that in the simulation of \cite{Haj-Hariri 1997}, and even bigger difference at the late stage. There are two reasons for the rapid decline in velocity:

\begin{figure}
\begin{minipage}[t]{0.48\linewidth}
\scalebox{1}[1]{\includegraphics[width=\linewidth]{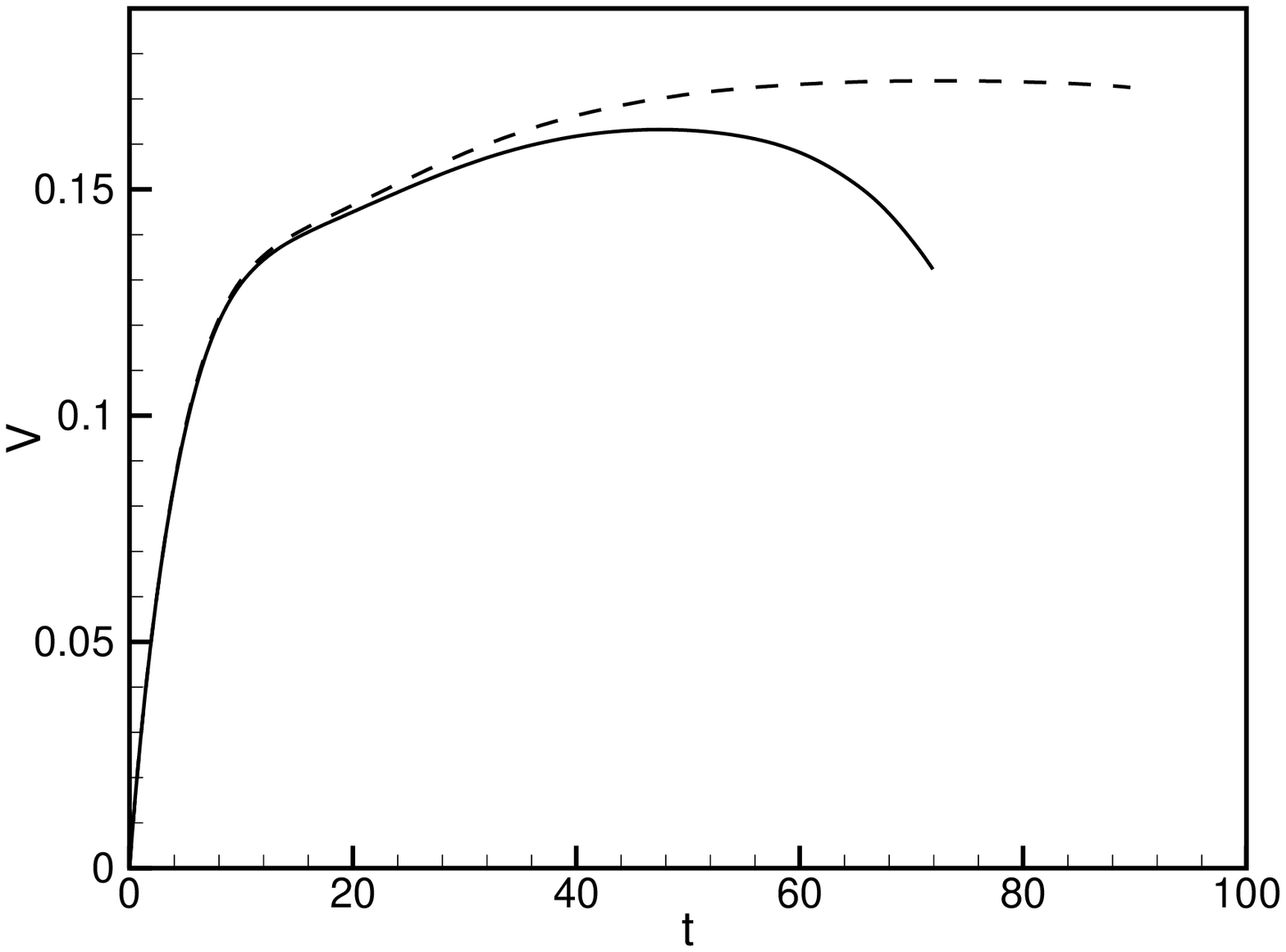}}
\caption{\label{fig:iso_time_distance_Ma}
Time evolution of drop velocity for $Re=Ma=50$, $\alpha=\lambda=\xi=0.5$, and $\gamma=0.25$.
$Ca=0.1$ (solid line); $Ca=0.05$ (dashed line).}
\end{minipage}
\hspace{0.1in}
\begin{minipage}[t]{0.48\linewidth}
\scalebox{1}[1]{\includegraphics[width=\linewidth]{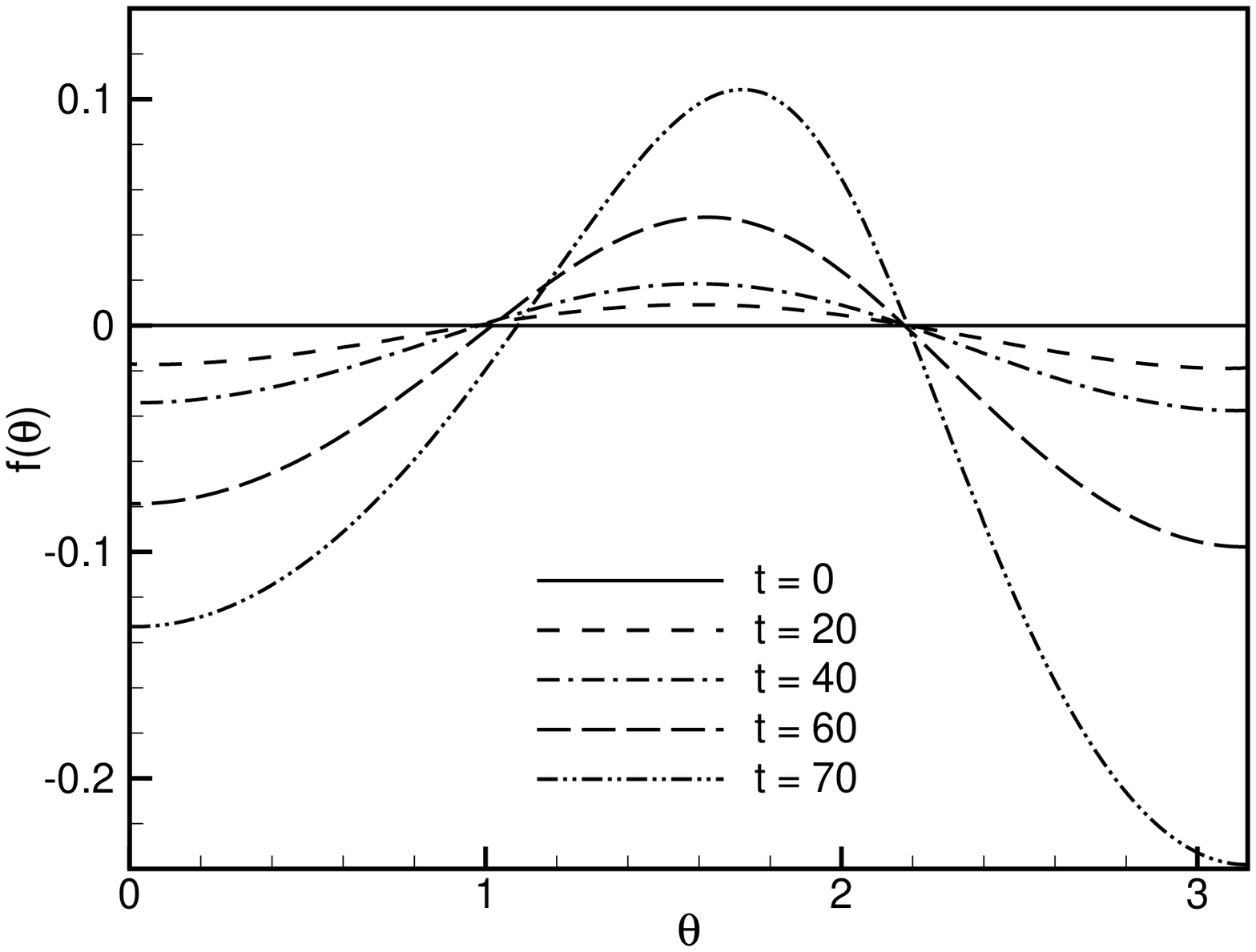}}
\caption{\label{fig:iso_time_distance_Ma}
Deviation of the drop profile from the sphere at indicated times for $Re=Ma=50$, $Ca=0.1$, $\alpha=\lambda=\xi=0.5$, and $\gamma=0.25$.}
\end{minipage}
\end{figure}

\begin{figure}
\begin{minipage}[t]{0.44\linewidth}
\scalebox{1}[1]{\includegraphics[width=\linewidth]{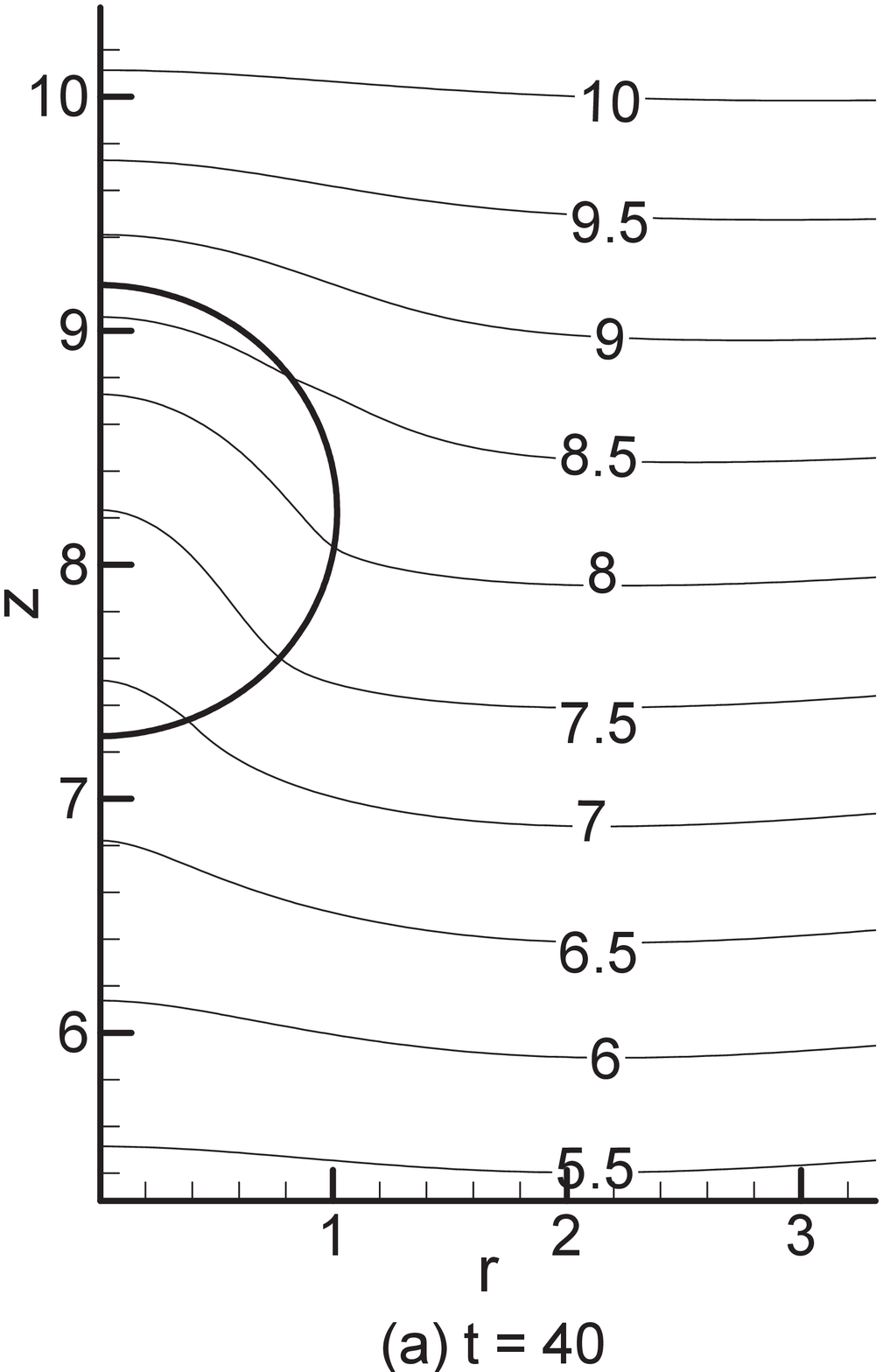}}
\end{minipage}
\hspace{0.6in}
\begin{minipage}[t]{0.44\linewidth}
\scalebox{1}[1]{\includegraphics[width=\linewidth]{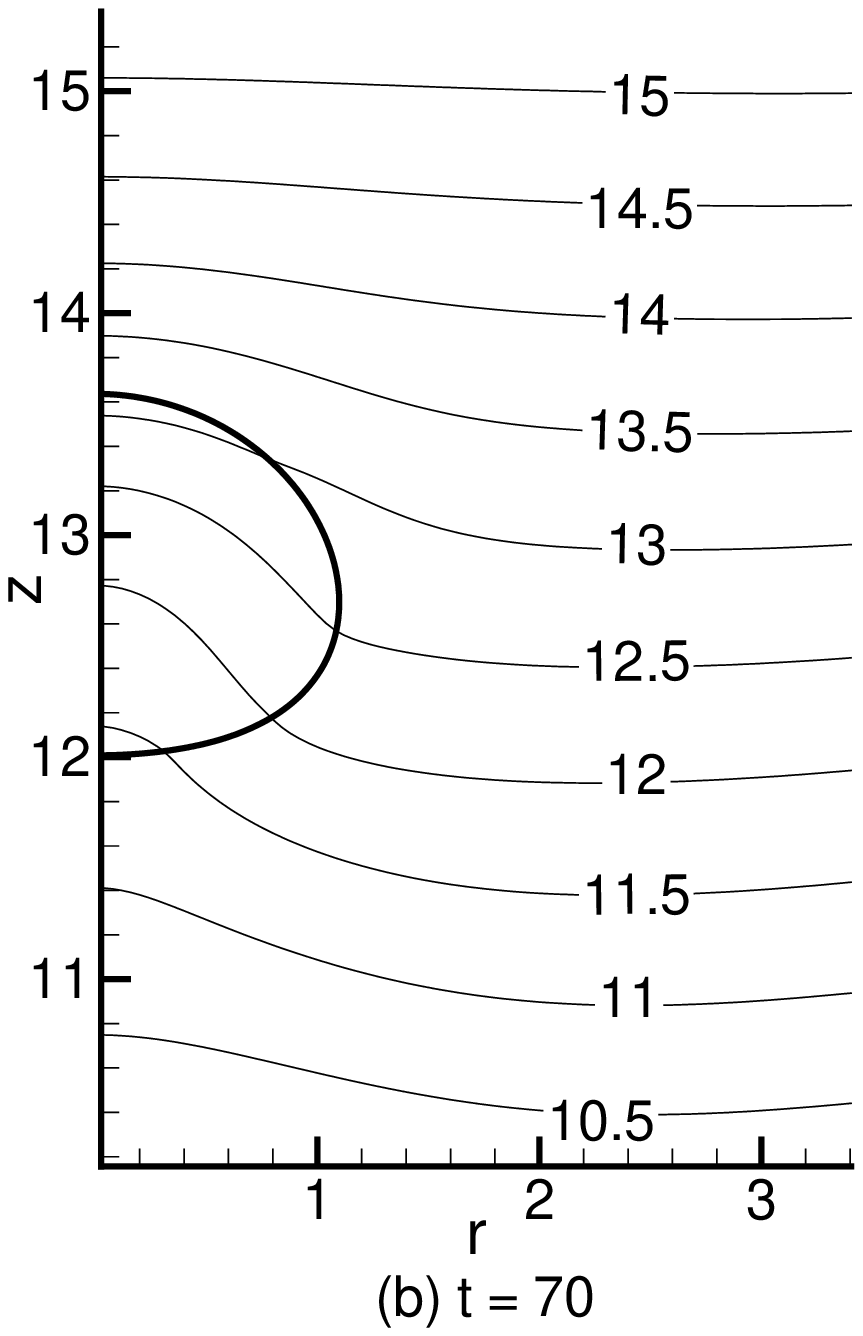}}
\end{minipage}
\caption{\label{fig:iso_time_distance_Ma}
Isotherms around the drop for $Re=Ma=50$, $Ca=0.1$, $\alpha=\lambda=\xi=0.5$, and $\gamma=0.25$. a) $t=40$; b) $t=70$.}
\end{figure}

\begin{figure}
\begin{minipage}[t]{0.45\linewidth}
\scalebox{1}[1]{\includegraphics[width=\linewidth]{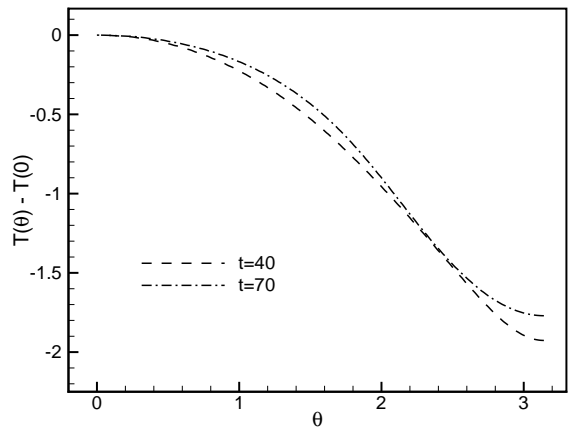}}
\end{minipage}
\caption{\label{fig:iso_time_distance_Ma}
The temperature distribution on the drop interface for $Re=Ma=50$, $Ca=0.1$, $\alpha=\lambda=\xi=0.5$, and $\gamma=0.25$ when $t=40$ and $t=70$.}
\end{figure}

\begin{enumerate}
\item Because the drop becomes flatter, the resistance on the drop is increased.
\item Because the distance between the front and rear stagnation points at the late stage is shorter than that in the beginning, their temperature difference also decreases (Fig. 7). As a result, the thermocapillary driving force on the drop becomes smaller.
\end{enumerate}

We also studied the influence of different $Re$ numbers, $Ma$ numbers, and $Ca$ numbers. Some typical results are shown in Figs. 8. With other parameters fixed, the migrating velocity tails off earlier for the larger Reynolds number.

With $Re=Ma=50$, $Ca=0.1$, and $\xi=0.5$, we studied the influence of $\alpha$, $\lambda$ and
$\gamma$. These three parameters play trivial roles in the transient motions of deformable
drops. With small $\alpha$, $\lambda$ or $\gamma$, drops migrate faster (see \cite{Nas 1995,
Yin 2008}) and need shorter time to reach the hot regions with near-zero interfacial tensions. As a result, the migration velocities tail off at earlier times.

To sum up, the influence of all other parameters except $\xi$ depends on when the drop migrates to the warm region with very low interfacial tension, in other words, the speed of the drop. The influence of various parameters on drop speeds has been discussed in detail in our earlier work \cite{Yin 2008,Gao 2007}.

\begin{figure}
\begin{minipage}[t]{0.49\linewidth}
\scalebox{1}[1]{\includegraphics[width=\linewidth]{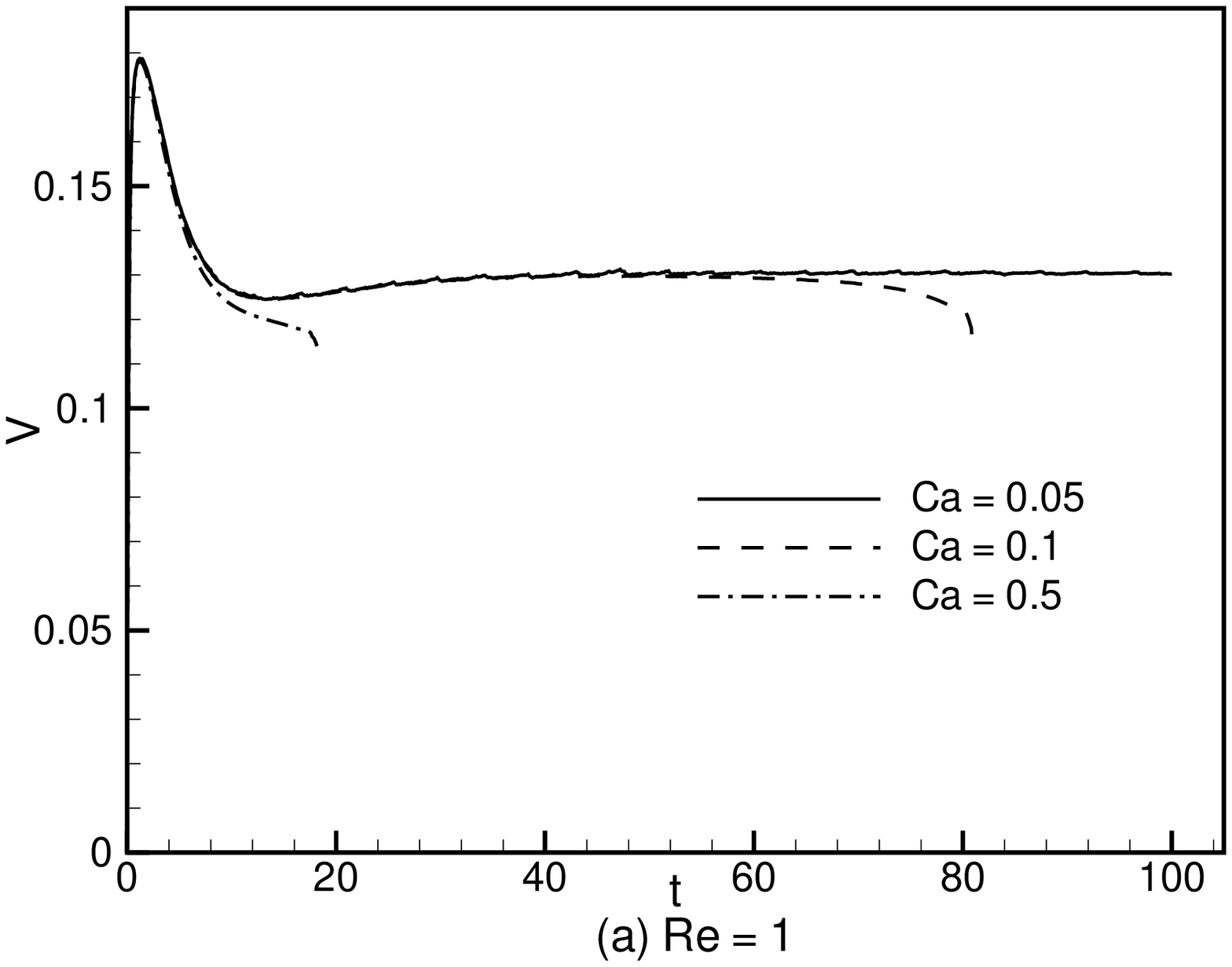}}
\end{minipage}
\begin{minipage}[t]{0.49\linewidth}
\scalebox{1}[1]{\includegraphics[width=\linewidth]{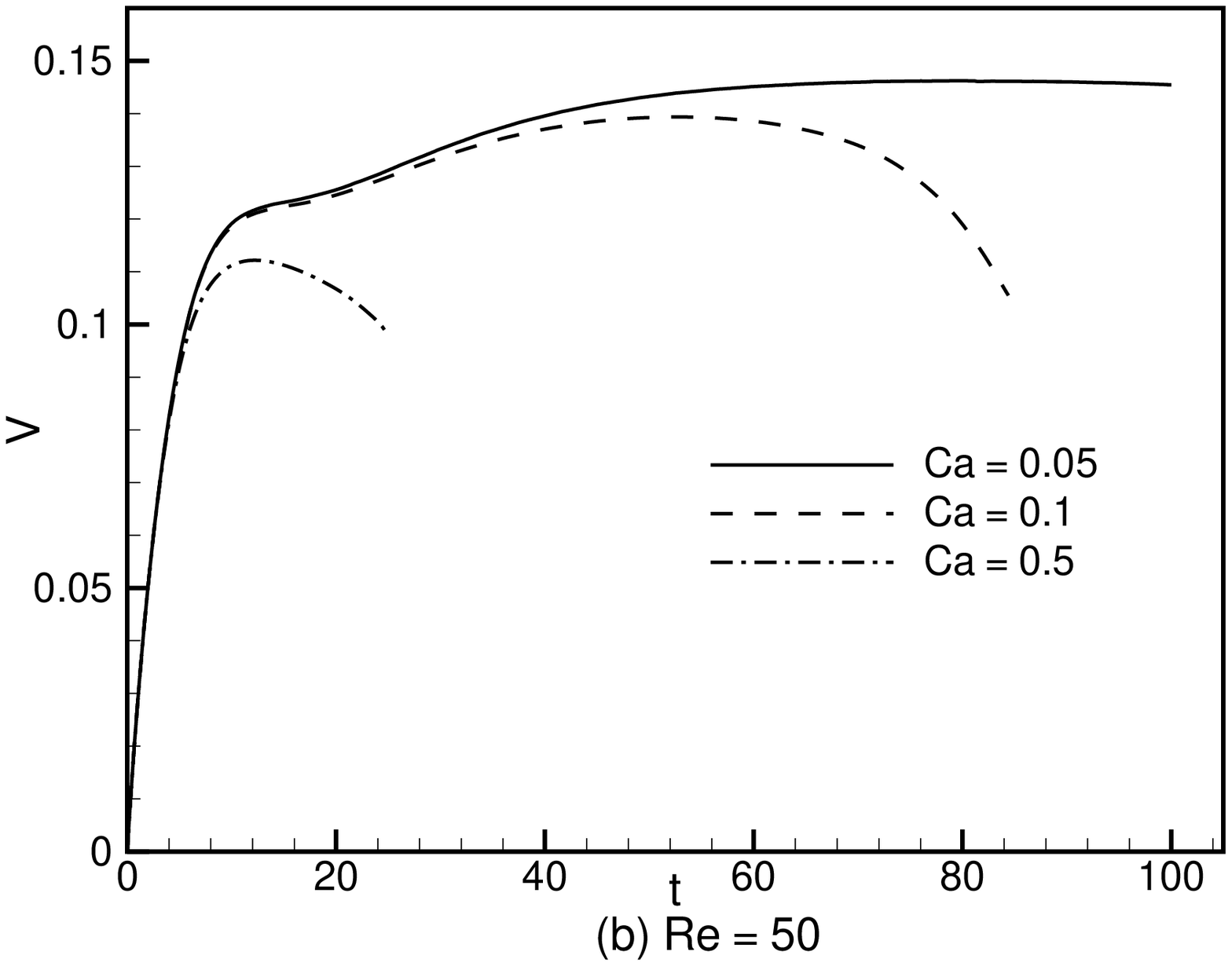}}
\end{minipage}
\caption{\label{fig:iso_time_distance_Ma}
Time evolutions of the drop velocity for different $Ca$ numbers. $Ma=100$, $\alpha=\lambda=\xi=0.5$, and $\gamma=0.25$. (a) $Re=1$; (b) $Re=50$.}
\end{figure}

\subsection{Influence of density ratio $\xi$}

\begin{figure}
\begin{minipage}[t]{0.8\linewidth}
\scalebox{1}[1]{\includegraphics[width=\linewidth]{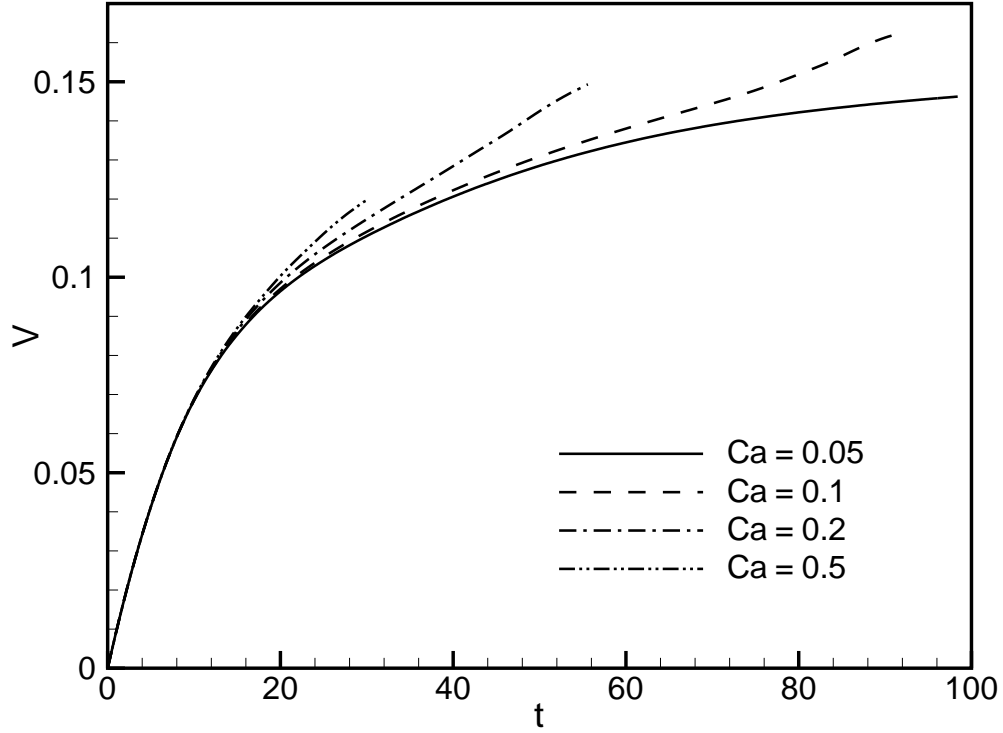}}
\end{minipage}
\caption{\label{fig:iso_time_distance_Ma}
Time evolutions of drop velocities for different capillary numbers. $Re=50$, $Ma=50$, $\alpha=\lambda=0.5$, $\gamma=0.25$, and $\xi=2$. }
\end{figure}

\begin{figure}
\begin{minipage}[t]{0.3\linewidth}
\scalebox{1}[1]{\includegraphics[width=\linewidth]{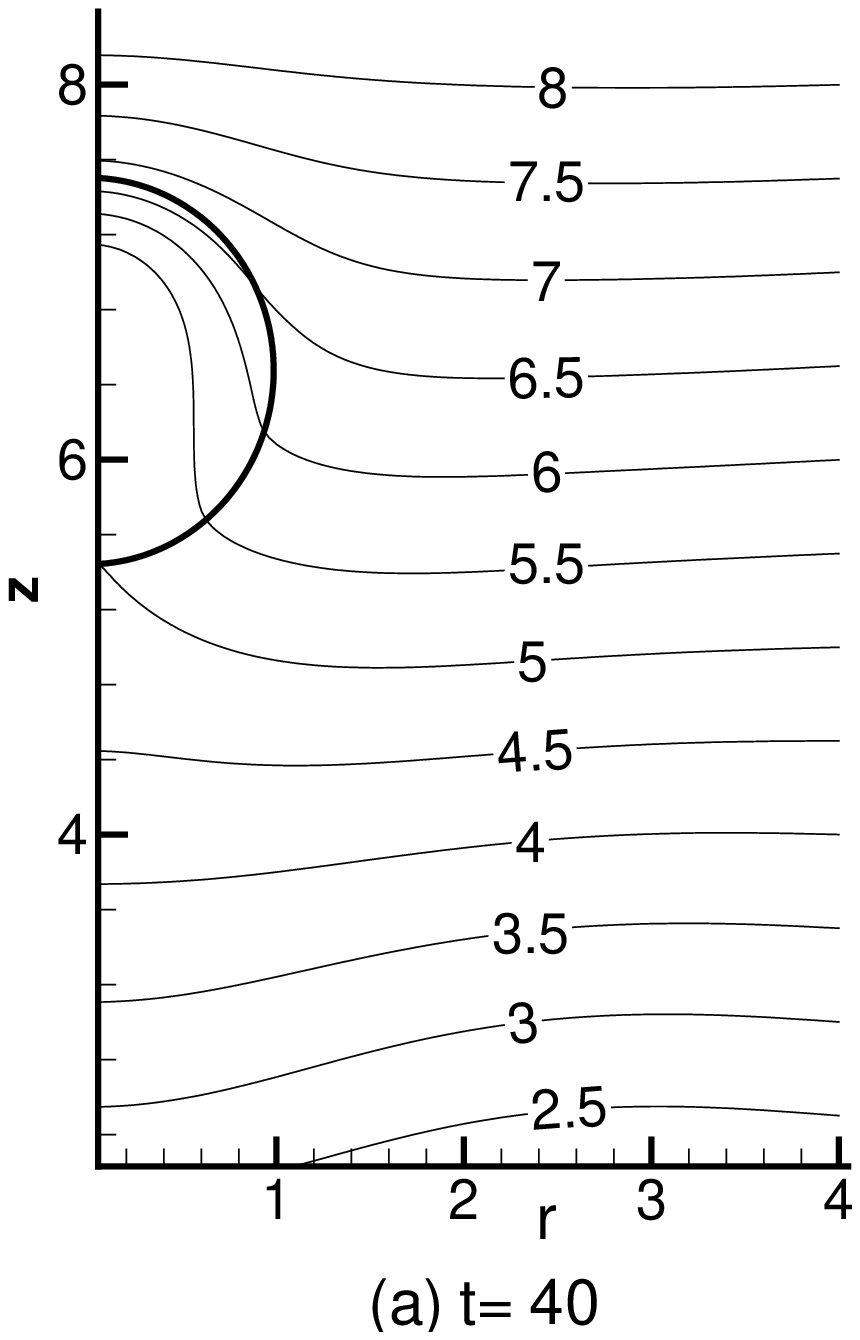}}
\end{minipage}
\hspace{1in}
\begin{minipage}[t]{0.3\linewidth}
\scalebox{1}[1]{\includegraphics[width=\linewidth]{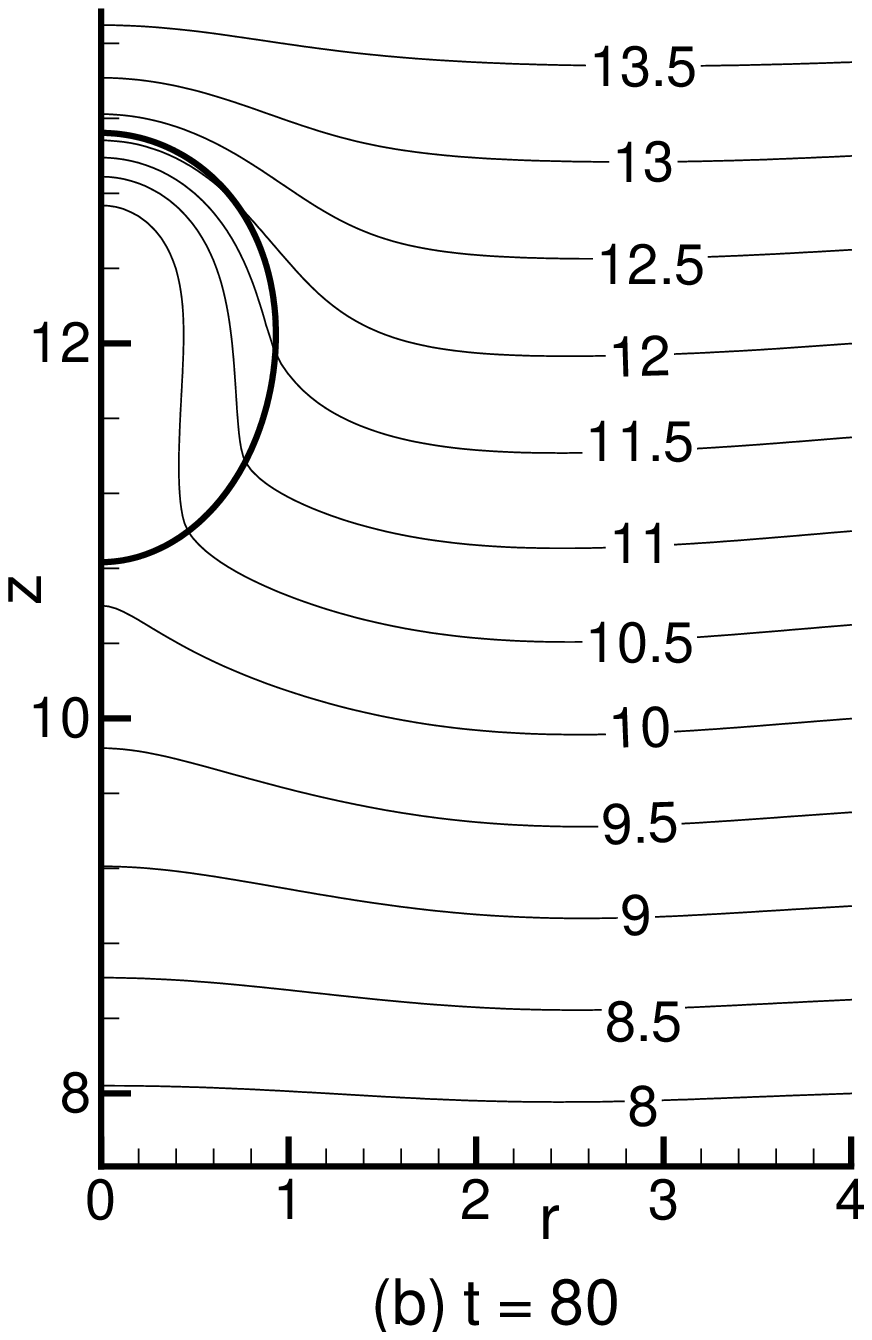}}
\end{minipage}
\caption{\label{fig:iso_time_distance_Ma}
Isotherms around the drop at $t=40$ and $t=80$. $Re=50$, $Ma=50$, $Ca=0.1$, $\alpha=\lambda=0.5$, $\gamma=0.25$, and $\xi=2$.}
\end{figure}

\begin{figure}
\begin{minipage}[t]{0.55\linewidth}
\scalebox{1}[1]{\includegraphics[width=\linewidth]{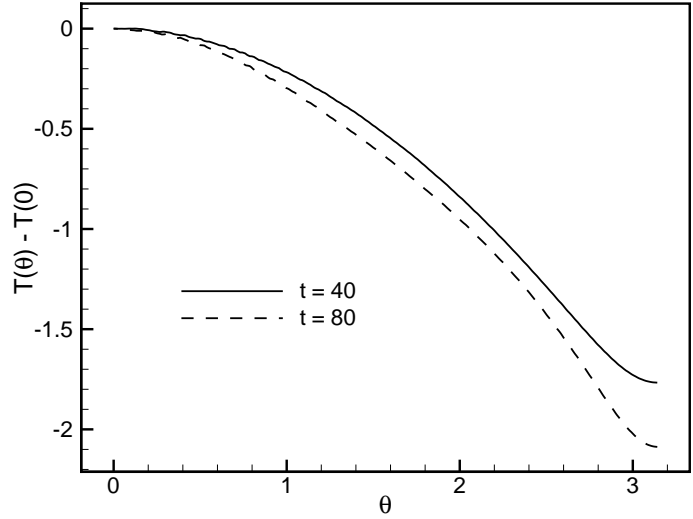}}
\end{minipage}
\caption{\label{fig:iso_time_distance_Ma}
Temperature distribution on the drop interface at $t=40$ and $t=80$.
$Re=50$, $Ma=50$, $Ca=0.1$, $\alpha=\lambda=0.5$, $\gamma=0.25$, and $\xi=2$.}
\end{figure}

\begin{figure}
\begin{minipage}[t]{0.45\linewidth}
\scalebox{1}[1]{\includegraphics[width=\linewidth]{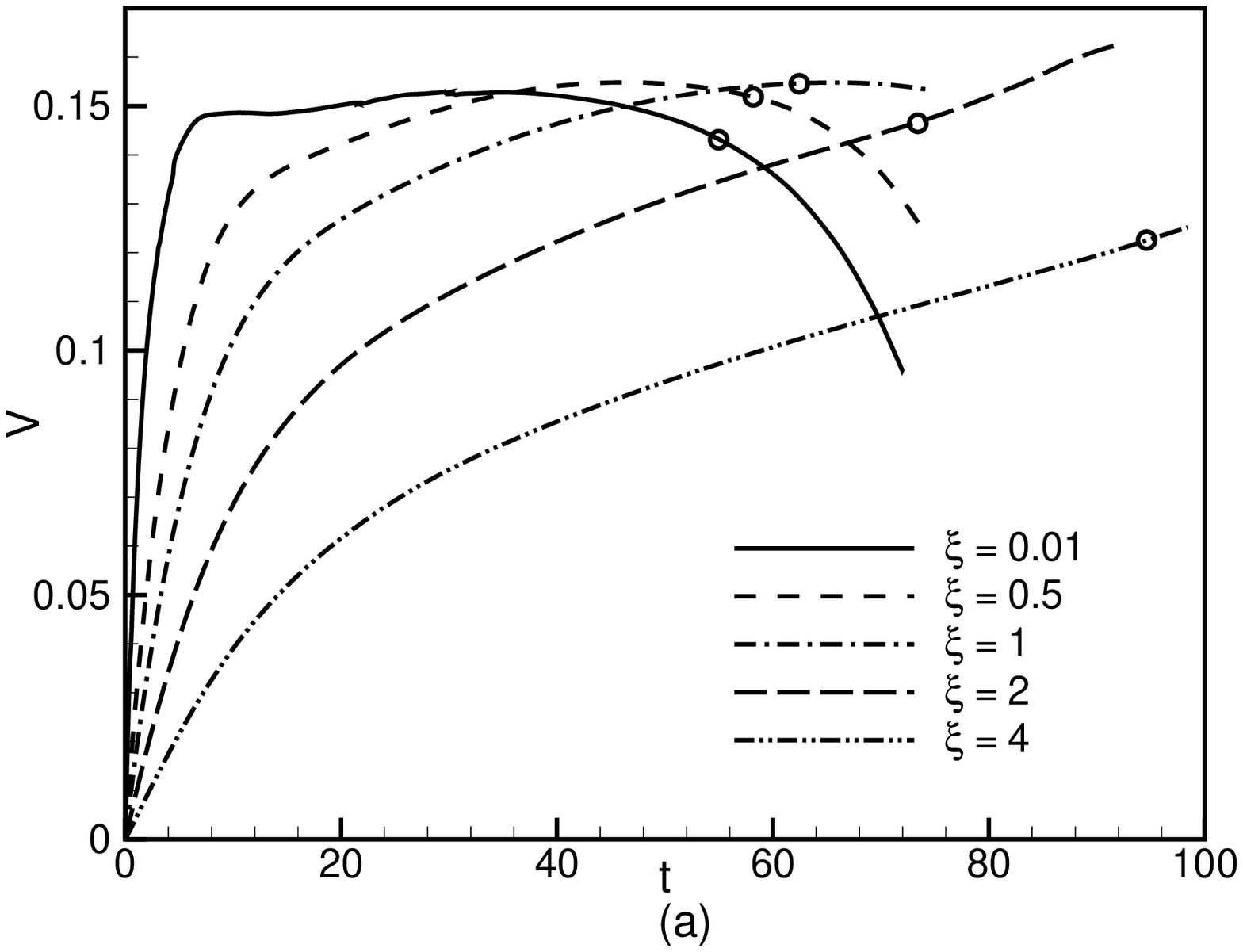}}
\end{minipage}
\hspace{0.5in}
\begin{minipage}[t]{0.45\linewidth}
\scalebox{1}[1]{\includegraphics[width=\linewidth]{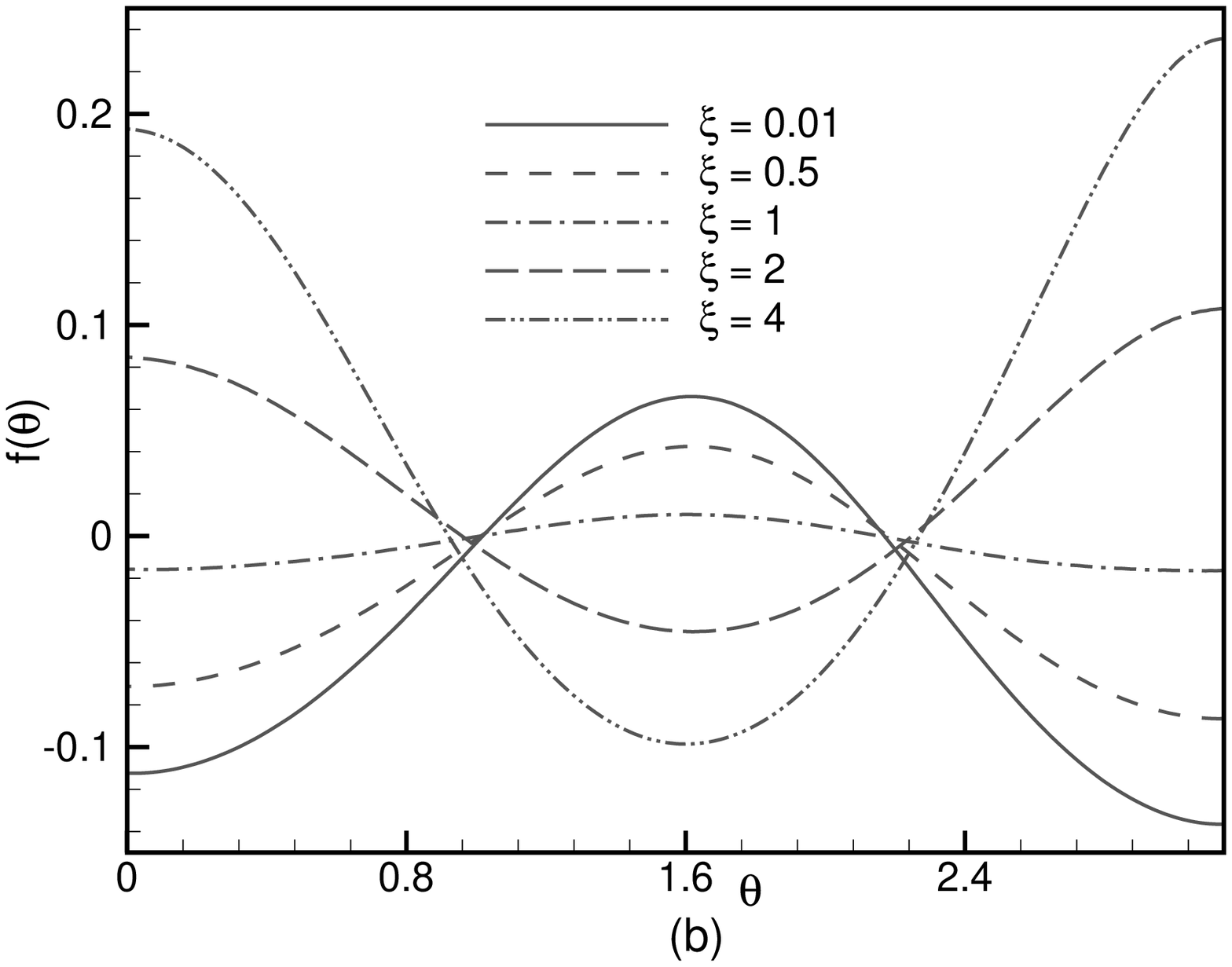}}
\end{minipage}
\caption{\label{fig:iso_time_distance_Ma}
(a) Time evolutions of drop velocities with different $\xi$'s, the circles on the velocity curves indicate the locations at $Ca_l=0.5$;  (b) deviations of drop profiles from spheres when $Ca_l=0.5$.
\\$Re=Ma=50$, $Ca=0.1$, $\alpha=\lambda=0.5$, and $\gamma=0.25$.  }
\end{figure}

\begin{figure}
\begin{minipage}[t]{0.8\linewidth}
\scalebox{1}[1]{\includegraphics[width=\linewidth]{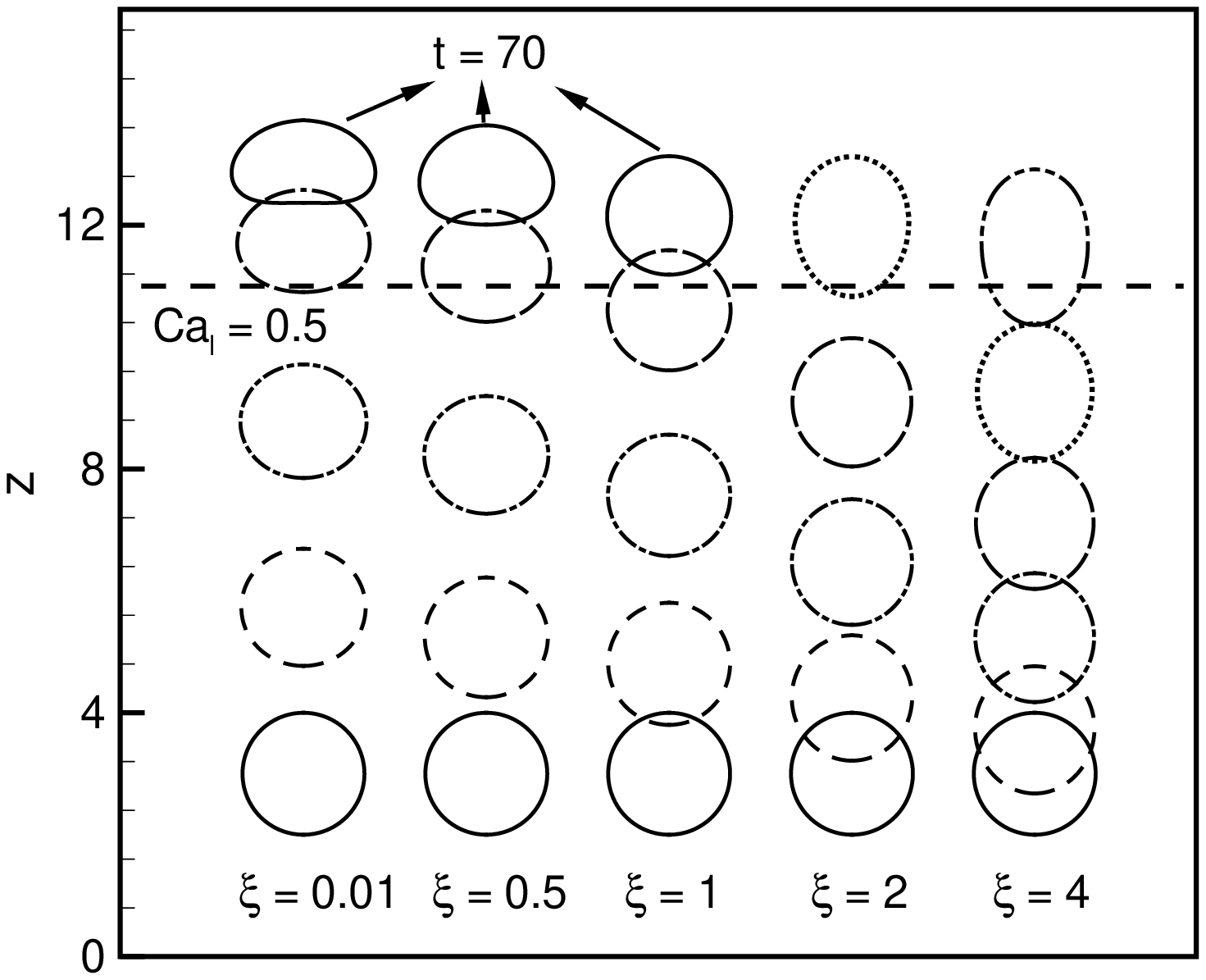}}
\end{minipage}
\caption{\label{fig:iso_time_distance_Ma}
Time evolutions of drop profiles for different $\xi$'s, where $Re=Ma=50$, $Ca=0.1$, $\alpha=\lambda=0.5$, and $\gamma=0.25$.  From the bottom to the top, the profiles in each column are corresponding to the time at $t=0, 20, 40, 60, 80, 100$, respectively.}
\end{figure}

In the last subsection, densities of drops are always smaller than those of bulk fluids, and we will adopt denser drops ($\xi=2$) in the following. Several $Ca$ numbers are studied, all other parameters are fixed to make the discussion simpler ($Re=Ma=50$, $\alpha=\lambda=0.5$, and $\gamma=0.25$). The time evolutions of velocities are plotted in Fig. 9, and the most obvious difference from Figs. 4\&8 is the continuous increasing trend of velocities. To explain this phenomenon, we focus on the case of $Ca=0.1$:
\begin{enumerate}
\item  At the late stage of the migration, the drop profile is elongated in the $z$ direction (Fig. 10(b)), and the temperature difference between the front and rear stagnation points of the drop becomes larger (Fig. 11). As a result, the thermocapillary driving force on the drop becomes larger.
\item  Because the drop becomes slender, the resistance on the drop decreases.
\end{enumerate}

To further discuss drop deformations, it is essential to introduce the local capillary number $Ca_l$ to indicate the local magnitude of interfacial tension, which is defined on the moving drop center ($z_c$):
\begin{equation}
Ca_l=\frac{Ca}{1-Ca(T_{c0}-T_0)}=\frac{Ca}{1-Ca(z_{c}-z_0)}, \label{eq19}
\end{equation}
where, $T_{c0}$ represents the initially temperature at $z_{c}$.
Time evolutions of drop velocities for different $\xi$'s are plotted in Fig. 12(a). To investigate the $\xi$ influence on deformation, we fixed $Ca_l$ at the value of $0.5$, when the drop starts to have an obviously different velocity from that of the non-deformable drop. In the current case ($Ca=0.1$), $Ca_l=0.5$ means that the drop migrates a distance of 8 (represented by circles on the curves of Fig. 12(a)).  It is found that the drop profile deforms to oblate/slender when $\xi$ is smaller/larger than unit. When the absolute value of $\xi-1$ becomes larger, the deformation is also larger (Fig. 12(b)). This is consistent in the analytical result (Eq. \ref{eq17}), except that we have larger deformation and $Re \& Ma$ numbers here.
Fig. 13 presents shape evolutions of drops with different $\xi$'s.

It should be noticed that only slight deformations on large bubbles ($\xi \sim 10^{-3}$) have been observed in previous space experiments, and no measurable deformations for those heavy drops ($\xi=1.98$) \cite{Hadland 1999}. The reason for the difference between experiments and simulations might be:
\begin{enumerate}
\item Bubbles/Drops in experiments do not migrate to areas with high temperatures, so their interface tensions are still large enough to hold perfect spherical shapes;

\item The assumption for the $\sigma (T)$'s linear dependence on temperature is not correct when $\sigma (T)\rightarrow 0$ in experiments (this is, however, the general assumption in most simulations in this field).
\end{enumerate}

\begin{figure}
\begin{minipage}[t]{0.7\linewidth}
\scalebox{1}[1]{\includegraphics[width=\linewidth]{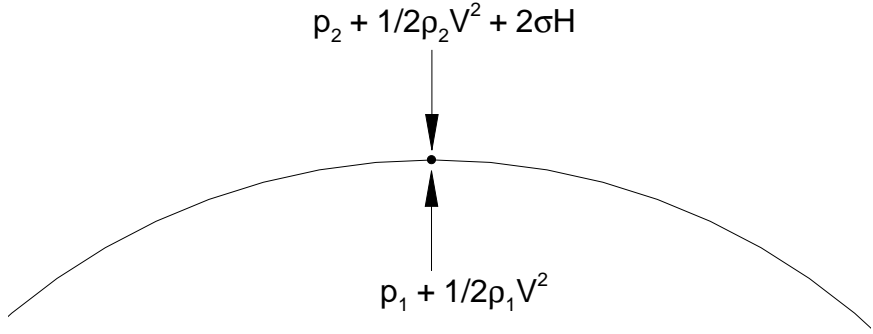}}
\end{minipage}
\caption{\label{fig:iso_time_distance_Ma}
Sketch of normal stress balance on the stagnation point of the drop.}
\end{figure}

To explain the relation between different $\xi$'s and drop-deforming orientations, a simplified physical explanation is introduced in the following. Only the front stagnation point is studied.

As shown in Fig. 14, the total pressures are $p_{2}+\frac{1}{2}\rho_2 V^2$ inside the drop and $p_{1}+\frac{1}{2}\rho_1 V^2$ outside the drop; here $V$ is the local fluid velocity, and $p_{2}$ and $p_{1}$ denote the static pressures at the two sides of the stagnation point, respectively.  The extra pressure caused by the interfacial tension is $2\sigma H$ (H is the average curvature). When the interfacial tension is large enough, the normal stress balance on the interface is dominated by the interfacial tension and the droplet keeps spherical. On the other hand, when the interfacial tension becomes small at the hot region (or, $Ca_l > 0.5$), the dynamic pressure becomes dominant:
\begin{description}
\item[if $\xi>1$,] the dynamic pressure inside the drop ($\frac{1}{2}\rho_2 V^2$) is larger than that outside ($\frac{1}{2}\rho_1 V^2$). Larger extra pressure ($2\sigma H$) is needed to balance the normal stress on the interface, so the drop is elongated in the axis direction to have a larger $H$.

\item[if $\xi<1$,] the dynamic pressure inside is lower than that outside. Smaller extra pressure is needed to balance the normal stress, so the drop is compacted in the axis direction to have a smaller $H$.

\end{description}

\section{Conclusions}

In this work, we found that the influence of deformations on drops is much more complicated than that on bubbles. When the drop density is smaller than the bulk fluid, the migration velocity will decrease with the increase of the deformation. When the drop density is larger than the bulk fluid, the migration velocity will increase with the increase of the deformation. With the assumption adopted in this paper, we believe that in the hot region the drop will always start to deform, and there is no way to have any constant migrating velocity. Hence, keeping $Ca$ and $Ca_l$ small throughout simulations is the only possible way for the drop to keep spherical and reach the steady migrating state.

All the discussions above are based on the assumption of the linear temperature dependence of the interfacial tension. More complicated temperature dependence will be used in the future work, and variations of physical properties with the temperature will also be considered.

\begin{acknowledgments}
This project is supported by the Knowledge Innovation Program of the Chinese Academy of Sciences, Grant No. KJCX2-YW-L08.
\end{acknowledgments}

\end{document}